\documentclass[conference]{IEEEtran}
\IEEEoverridecommandlockouts
\usepackage{cite}
\usepackage{amsmath,amssymb,amsfonts}
\usepackage{graphicx}
\usepackage{textcomp}
\usepackage{xcolor}
\usepackage{algorithm,algorithmicx,algpseudocode}
\usepackage[]{footmisc}
\usepackage[caption=false,font=footnotesize,subrefformat=parens]{subfig}
\def\BibTeX{{\rm B\kern-.05em{\sc i\kern-.025em b}\kern-.08em
    T\kern-.1667em\lower.7ex\hbox{E}\kern-.125emX}} 

\newcommand{\norm}[1]{\left\lVert#1\right\rVert}

\DeclareMathOperator*{\argmin}{arg\,min}
\begin{document}

\title{Hybrid Beamformer Codebook Design and Ordering for Compressive mmWave Channel Estimation\\
}

\author{\IEEEauthorblockN{
Junmo Sung and Brian L. Evans}
\IEEEauthorblockA{\textit{Wireless Networking and Communications Group},
\textit{The University of Texas at Austin}
Austin, TX USA \\
junmo.sung@utexas.edu, bevans@ece.utexas.edu}
}

\maketitle

\begin{abstract}
In millimeter wave (mmWave) communication systems, beamforming with large antenna arrays is critical to overcome high path losses. Separating all-digital beamforming into analog and digital stages can provide the large reduction in power consumption and small loss in spectral efficiency needed for practical implementations. Developing algorithms with this favorable tradeoff is challenging due to the additional degrees of freedom in the analog stage and its accompanying hardware constraints.  In hybrid beamforming systems, for example, channel estimation algorithms do not directly observe the channels, face a high channel count, and operate at low SNR before transmit-receive beam alignment.  Since mmWave channels are sparse in time and beam domains, many compressed sensing (CS) channel estimation algorithms have been developed that randomly configure the analog beamformers, digital beamformers, and/or pilot symbols.  In this paper, we propose to design deterministic beamformers and pilot symbols for open-loop channel estimation.  We use CS approaches that rely on low coherence for their recovery guarantees, and hence seek to minimize the mutual coherence of the compressed sensing matrix.  We also propose a precoder column ordering to design the pilot symbols.  Simulation results show that our beamformer designs reduce channel estimation error over competing methods.
\end{abstract}


\section{Introduction}
Hybrid analog and digital beamforming architectures in millimeter wave (mmWave) communication systems have drawn a great amount of attention for multiple reasons. They can practically maintain achievable spectral efficiency as with all-digital MIMO architectures due to the sparse nature of mmWave channels \cite{sohrabi2016stsp,ayach2014twc}. Reducing the number of RF chains leads to almost proportional power consumption reduction. Hybrid beamforming architectures, however, demand even more complicated signal processing \cite{heath2016jstsp} because an analog stage has fewer degrees of freedom compared with an all-digital MIMO architecture; e.g., phase shifters are constrained to have a discrete phase on the unit circle. Channel estimation is not an exception in this signal processing complication. 

MmWave channel measurement campaigns have revealed that the channels are sparse in both time and angular dimensions \cite{rapp2015tcom,rapp2013tap}. The sparsity finds compressed sensing (CS) algorithms suitable for mmWave channel estimation. For phase shifter based hybrid beamforming architecture, the adaptive CS was proposed in \cite{Alkh2014stsp} to narrow beams by iteratively adapting precoders and combiners. Another popular approach found in many publications is to formulate a sparse channel estimation problem and apply well-known or modified CS algorithms \cite{Venu2017,rf2018twc,MR2016access,park2016asilomar,gao2017twc}. For example, orthogonal matching pursuit (OMP) or its variants are used in \cite{Venu2017,rf2018twc}, and \cite{MR2016access} compares different channel estimation algorithms that include OMP as a representative of CS algorithms. The adaptive dictionary generation algorithm proposed in \cite{xiao2019access} is useful to deal with common issues regarding the dimension deficiency.

The random configuration for the phase shifters has been widely used in those literatures since the generated sensing matrices are incoherent and satisfy the restricted isometry property condition with high probability. In contrast with the random configuration, a deterministic codebook and pilot design was considered in \cite{lee2016tcom}. The approach taken in \cite{lee2016tcom} is to obtain codebooks that provide the minimal total coherence (MTC) of the sensing matrix. However, it still contains randomness in beamformer column permutation. 

In this paper, we propose a completely deterministic beamformer codebook and pilot design method for CS based open-loop narrowband mmWave channel estimation. In practice, receivers should know about pilots and precoding schemes in order to properly perform channel estimation even with the random codebook. Therefore we design codebooks and pilots that can minimize mutual coherence of the resulting sensing matrix, which is of importance to CS algorithms. As suggested in \cite{lee2016tcom}, random beamformer column permutation plays a critical role to reduce measurement time and instances. We propose a greedy algorithm to find the best column permutation of the obtained RF beamformer codebook. Simulation results show that codebooks obtained by the proposed method outperforms the random codebook, the MTC codebook, and the adaptive CS in practical situations. 

\begin{figure*}[t!]
	\centering
	\includegraphics[width=14.5cm]{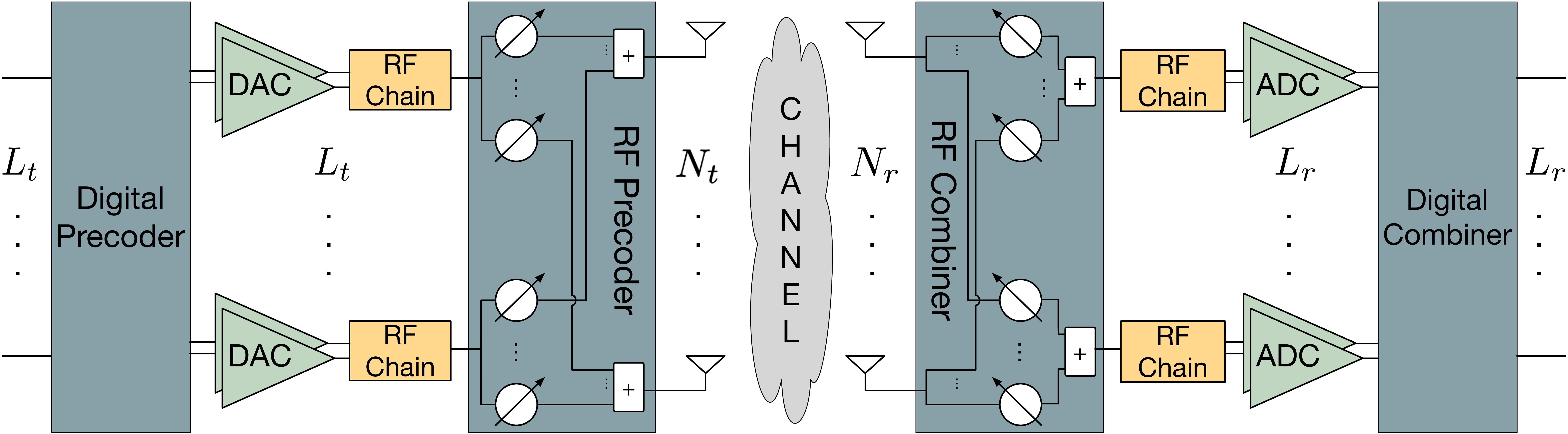}
	\caption{System block diagram of the phase shifter based hybrid beamforming architecture}
	\label{fig:system}
\end{figure*}

\section{System Model} 
\label{sec:system_model}
In the downlink, we assume a single base station (BS)
and channel estimation is performed by each user. Both the BS and user equipment (UE) are equipped with the fully-connected phase shifter based hybrid beamforming architecture as shown in Fig~\ref{fig:system}. The BS has $N_t$ transmit antennas and $L_t$ transmit RF chains, and the UE has $N_r$ receive antennas and $L_r$ receive RF chains. The number of streams is assumed to be equal to $L_t$. Assuming the channel is narrowband, the received signal after RF and baseband combining at the time instance $m$ can be expressed as
\begin{align}
	\label{eq:y_m}
	\mathbf{y}_m = \sqrt{\rho} \mathbf{W}_{m}^{\mathsf{H}} \mathbf{H} \mathbf{F}_{m} \mathbf{x}_{m} + \mathbf{W}_{m}^{\mathsf{H}} \mathbf{n}_{m} \in \mathbb{C}^{L_r}, 
\end{align}
where $\rho$ is the average transmit power, $\mathbf{W}_{m} \in \mathbb{C}^{N_r \times L_r}$ and $\mathbf{F}_{m} \in \mathbb{C}^{N_t \times L_t}$ denote the combiner and precoder matrices, respectively, $\mathbf{x}_{m} \in \mathbb{C}^{L_t}$ denotes the pilot symbol vector, $\mathbf{H} \in \mathbb{C}^{N_r \times N_t}$ is the channel matrix, and $\mathbf{n}_{m} \in \mathbb{C}^{N_r} \sim \mathcal{CN}(0, \sigma_n^2 \mathbf{I})$ denotes the additive noise. $(\cdot)^{\mathsf{H}}$ denotes the conjugate transpose. Both the precoder and combiner matrices are a product of RF and baseband ones, i.e., 
$\mathbf{F}_{m} = \mathbf{F}_{RF,m} \mathbf{F}_{BB,m}$ and $\mathbf{W}_{m} = \mathbf{W}_{RF,m} \mathbf{W}_{BB,m}$. By vectorizing the right hand side in \eqref{eq:y_m}, it can be rewritten as
\begin{align}
	\mathbf{y}_{m} 
	&= 
		\sqrt{\rho} \left(
			\mathbf{s}_{m}^{\mathsf{T}} \otimes \mathbf{W}_{m}^{\mathsf{H}}
		\right)
		\mathrm{vec}(\mathbf{H}) + \mathbf{v}_{m}, \nonumber \\
	&=
		\sqrt{\rho} \mathbf{\Phi}_m \mathrm{vec}(\mathbf{H}) + \mathbf{v}_{m}, \nonumber
\end{align}
where $\mathbf{s}_{m} = \mathbf{F}_{m} \mathbf{x}_{m}$, $\mathbf{v}_{m} = \mathbf{W}_{m}^{\mathsf{H}} \mathbf{n}_{m} $, $\mathbf{\Phi}_m = \mathbf{s}_{m}^{\mathsf{T}} \otimes \mathbf{W}_{m}^{\mathsf{H}}$, and $(\cdot)^{\mathsf{T}}$ and $\otimes$ denote the matrix transpose and the Kronecker product. By stacking $M$ instances of the received signal vectors, we can obtain
\begin{align}
	\label{eq:y}
	\mathbf{y} = \sqrt{\rho} \mathbf{\Phi} \mathrm{vec}(\mathbf{H}) + \mathbf{v} \in \mathbb{C}^{M L_r}, 
\end{align}
where 
$\mathbf{y} = [\mathbf{y}_{1}^{\mathsf{T}}, \mathbf{y}_{2}^{\mathsf{T}}, \ldots, \mathbf{y}_{M}^{\mathsf{T}}]^{\mathsf{T}}$ is the total received signal vector, 
$\mathbf{\Phi} = [\mathbf{\Phi}_{1}^{\mathsf{T}}, \mathbf{\Phi}_{2}^{\mathsf{T}}, \ldots, \mathbf{\Phi}_{M}^{\mathsf{T}}]^{\mathsf{T}}$ is the sensing matrix, 
and 
$\mathbf{v} = [\mathbf{v}_{1}^{\mathsf{T}}, \mathbf{v}_{2}^{\mathsf{T}}, \ldots, \mathbf{v}_{M}^{\mathsf{T}}]^{\mathsf{T}}$. 

For the narrowband channel, we adopt the geometric channel model. Assuming $N_p$ clusters constitute the channel, the channel matrix can be given as
\begin{align}
	\label{eq:H_vector_form}
	\mathbf{H} = \sqrt{\frac{N_t N_r}{N_p}} \sum_{l = 0}^{N_p-1} 
		\alpha_{l} \mathbf{a}_{r}(\theta_l) \mathbf{a}_{t}^{\mathsf{H}}(\vartheta_l) \in \mathbb{C}^{N_r \times N_t},
\end{align}
where $\alpha_l \sim \mathcal{CN}(0,\sigma_\alpha^2)$ is the complex channel gain, $\mathbf{a}_{t}(\cdot)$ and $\mathbf{a}_{r}(\cdot)$ are, respectively, the transmit and receive array response vectors evaluated at the angles. 
$\theta$ and $\vartheta$ are the angle of arrival (AoA) and departure (AoD).
In this paper, the employed antenna arrays are assumed to be uniform linear array with a half wavelength spacing. The transmit and receive array response vectors are then given as
\begin{align}
	\mathbf{a}_{t}(\vartheta) = \sqrt{\frac{1}{N_t}}
		[1, e^{-j\pi \cos\vartheta}, \ldots, e^{-j\pi \cos(N_t-1)\vartheta}]^{\mathsf{T}} \in \mathbb{C}^{N_t}, \nonumber 
	\\
	\mathbf{a}_{r}(\theta) = \sqrt{\frac{1}{N_r}}
		[1, e^{-j\pi \cos\theta}, \ldots, e^{-j\pi \cos(N_r-1)\theta}]^{\mathsf{T}} \in \mathbb{C}^{N_r}, \nonumber 
\end{align}
and the corresponding transmit and receive array response matrices can be constructed as
\begin{align}
	\mathbf{A}_{t} = [\mathbf{a}_{t}(\vartheta_0), \mathbf{a}_{t}(\vartheta_1), \ldots, \mathbf{a}_{t}(\vartheta_{N_p-1})] \in \mathbb{C}^{N_t \times N_p}, \nonumber
	\\
	\mathbf{A}_{r} = [\mathbf{a}_{r}(\theta_0), \mathbf{a}_{r}(\theta_1), \ldots, \mathbf{a}_{r}(\theta_{N_p-1})] \in \mathbb{C}^{N_r \times N_p}. \nonumber
\end{align}
With the transmit and receive array response matrices, \eqref{eq:H_vector_form} can be rewritten as
\begin{align}
	\label{eq:H_matrix_form}
	\mathbf{H} = \mathbf{A}_{r} \mathbf{H}_{d} \mathbf{A}_{t}^{\mathsf{H}},
\end{align}
where $\mathbf{H}_d \in \mathbb{C}^{N_p \times N_p}$ is a diagonal matrix with the scaled channel gains on its diagonal. 

Now we define the angle dictionary matrix for sparse formulation. The transmit and receive angle dictionary matrix, $\bar{\mathbf{A}}_t$, is defined as
\begin{align}
	\bar{\mathbf{A}}_{t} = \sqrt{\frac{1}{N_t}} \left[
		\mathbf{a}_t(\nu_0), \mathbf{a}_t(\nu_1), \ldots, \mathbf{a}_t(\nu_{G_t-1}) 
	\right]  \in \mathbb{C}^{N_t \times G_t}, \nonumber \\
\end{align}
where $G_t$ is the size of transmit angle grid, and $\nu_{n} \in \Theta = \{ \nu_n=\arccos\frac{\phi_n}{\pi} \ | \ \phi_n = \frac{n}{G_t} - \frac{1}{2}, n=0,1, \ldots, G_t-1 \}$ where $G_t \geq N_t$. It it worth noting that $\bar{\mathbf{A}}_t \bar{\mathbf{A}}_t^{\mathsf{H}} = \frac{G}{N_t} \mathbf{I}_{N_t}$ where $\mathbf{I}_{N_t}$ is the $N_t \times N_t$ identity matrix. The receive angle dictionary matrix, $\bar{\mathbf{A}}_r$, is also similarly defined with $G_r (\geq N_r)$. Then the channel matrix in \eqref{eq:H_matrix_form} can be rewritten as
\begin{align}
	\label{eq:H}
	\mathbf{H} = \bar{\mathbf{A}}_{r} \bar{\mathbf{H}}_{d} \bar{\mathbf{A}}_{t}^{\mathsf{H}},
\end{align}
where $\bar{\mathbf{H}}_{d} \in \mathbb{C}^{G_r \times G_t}$ is the channel gain matrix which needs not be a diagonal matrix. Ignoring the grid quantization errors, $\mathrm{vec}(\bar{\mathbf{H}}_{d})$ is an $N_p$-sparse vector that will be estimated. 

By plugging \eqref{eq:H} into \eqref{eq:y}, we can get
\begin{align}
	\mathbf{y} 
	&= 
		\sqrt{\rho} \mathbf{\Phi} ( \bar{\mathbf{A}}_{t}^* \otimes \bar{\mathbf{A}}_{r}) \mathrm{vec}(\bar{\mathbf{H}}_d) + \mathbf{v}
		\nonumber
		\\
	&= 
		\sqrt{\rho} \mathbf{\Phi} \mathbf{\Psi} \mathbf{h} + \mathbf{v},
		\nonumber
\end{align}
where $\mathbf{\Psi} = \bar{\mathbf{A}}_{t}^* \otimes \bar{\mathbf{A}}_{r} \in \mathbb{C}^{N_t N_r \times G_t G_r}$ is the dictionary matrix, and $\mathbf{h} = \mathrm{vec}(\bar{\mathbf{H}}_d) \in \mathbb{C}^{G_t G_r}$ is the sparse channel gain vector. The product of the sensing matrix and the dictionary matrix is called the equivalent sensing matrix, i.e., $\mathbf{\Phi} \mathbf{\Psi}$. We assume that the channel gain vector is estimated by using CS algorithms. 

\section{Coherence Minimizing Codebook} 
\label{sec:coherence_minimizing_codebook}
Recovery guarantees of CS algorithms can be assessed by mutual coherence of the equivalent sensing matrix, and the mutual coherence, $\mu$, is defined by
\begin{align}
	\mu(\mathbf{A}) = \max_{i \neq j} \frac{|\mathbf{a}_i^{\mathsf{H}} \mathbf{a}_j|}{\norm{\mathbf{a}_i}_2 \norm{\mathbf{a}_j}_2 }
	=
	\max_{i \neq j} \frac{|(\mathbf{A}^{\mathsf{H}} \mathbf{A})_{ij}|}{\norm{\mathbf{a}_i}_2 \norm{\mathbf{a}_j}_2 }, 
	\nonumber
\end{align}
where $\mathbf{a}_i$ denotes the $i$-th column vector in the matrix $\mathbf{A}$, and $\mathbf{A}_{ij}$ denotes the element in the $i$-the row and $j$-th column of the matrix $\mathbf{A}$. 
Therefore mutual coherence minimization can be achieved by minimizing all off-diagonal elements of $\mathbf{A}^{\mathsf{H}} \mathbf{A}$, i.e., $\min \norm{ \mathbf{A}^{\mathsf{H}} \mathbf{A} - \mathbf{I} }_F^2$. Plugging the equivalent sensing matrix, the optimization problem is given by
\begin{align}
	\label{eq:totmut_min}
	\min \norm{ \mathbf{\Psi}^{\mathsf{H}} \mathbf{\Phi}^{\mathsf{H}} \mathbf{\Phi} \mathbf{\Psi} - \mathbf{I} }_F^2.
\end{align}
The objective function in \eqref{eq:totmut_min} can be simplified as
\begin{align}
	\norm{\mathbf{\Psi}^{\mathsf{H}} \mathbf{\Phi}^{\mathsf{H}} \mathbf{\Phi} \mathbf{\Psi} - \mathbf{I}}_F^2
	&=
		\norm{k \mathbf{\Phi}^{\mathsf{H}} \mathbf{\Phi} - \mathbf{I}  }_F^2
		\nonumber
		\\
	&=
		\norm{
			k \sum_{m=1}^{M} \mathbf{\Phi}_m^{\mathsf{H}} \mathbf{\Phi}_m - \mathbf{I}
		}_F^2,
		\nonumber
\end{align}
where 
the first equality holds due to the fact that $\mathbf{\Psi} \mathbf{\Psi}^{\mathsf{H}}=k \mathbf{I}$ and
$k = \frac{G_t G_r}{N_t N_r}$ can be ignored since it scales all columns of the equivalent sensing matrix and does not affect mutual coherence. Our goal, thus, is to find the sets of the pilots, precoders and combiners that minimize the objective function. Considering the fact that $\mathbf{\Phi}_m$ is $\mathbf{s}_{m}^{\mathsf{T}} \otimes \mathbf{W}_{m}^{\mathsf{H}}$, the summation in the objective function can also be written as
\begin{align}
	\sum \mathbf{\Phi}_m^{\mathsf{H}} \mathbf{\Phi}_m
	= \sum \mathbf{s}_m^* \mathbf{s}_m^{\mathsf{T}} \otimes \mathbf{W}_m \mathbf{W}_m^{\mathsf{H}}.
	\nonumber
\end{align}
The above equation can readily be made the identity matrix if (i) the summation can be distributed over the Kronecker product and (ii) $M$ is large enough. For illustration, let $M$ be $M=M_t M_r$ 
where $M_t$ and $M_r$ denote the number of configurations of transmitter and receiver, 
then we have
\begin{align}
	\label{eq:ssWW}
	\sum_{m}^{M} \mathbf{s}_m^* \mathbf{s}_m^{\mathsf{T}} \otimes \mathbf{W}_m \mathbf{W}_m^{\mathsf{H}}
	&=
		\sum_{m_t}^{M_t} \sum_{m_r}^{M_r} \mathbf{s}_{m_t}^* \mathbf{s}_{m_t}^{\mathsf{T}} \otimes \mathbf{W}_{m_r} \mathbf{W}_{m_r}^{\mathsf{H}}
		\nonumber
		\\
	&= \sum_{m_t}^{M_t} \mathbf{s}_{m_t}^* \mathbf{s}_{m_t}^{\mathsf{T}} \otimes
		\sum_{m_r}^{M_r} \mathbf{W}_{m_r} \mathbf{W}_{m_r}^{\mathsf{H}}.
\end{align}
Since $\mathbf{s}_{m_t}^* \mathbf{s}_{m_t}^{\mathsf{T}}$ is a rank one matrix, and $\mathbf{W}_{m_r} \mathbf{W}_{m_r}^{\mathsf{H}}$ can be up to a rank $L_r$ matrix, $M_t$ and $M_r$ must be at least $N_t$ and $\frac{N_r}{L_r}$ in order to make both $\sum \mathbf{s}_{m_t}^* \mathbf{s}_{m_t}^{\mathsf{T}}$ and $\sum \mathbf{W}_{m_r} \mathbf{W}_{m_r}^{\mathsf{H}}$ become a full rank matrix. Being full rank matrices is a critical requirement for it to be the identity matrix. Then the solutions for making both terms the identity matrix can easily be found. We now set this as the baseline, and start to decrease $M$.

As $M$ is a product of $M_t$ and $M_r$, either or both can be decreased to have a lower $M$ value. A BS normally has far more antennas than UEs do ($N_t \gg \frac{N_r}{L_r}$), we target $M_t$ for $M$ reduction. Assuming $\sum \mathbf{W}_{m_r} \mathbf{W}_{m_r}^{\mathsf{H}} = \mathbf{I}$, we focus on $\sum \mathbf{s}_{m_t}^* \mathbf{s}_{m_t}^{\mathsf{T}}$. It can be rewritten as
\begin{align}
	\label{eq:sum_ss}
	\sum_{m_t}^{M_t} \mathbf{s}_{m_t}^* \mathbf{s}_{m_t}^{\mathsf{T}} 
	&=
		\sum_{m_t}^{M_t} \mathbf{F}_{m_t}^* \mathbf{x}_{m_t}^* \mathbf{x}_{m_t}^{\mathsf{T}} \mathbf{F}_{m_t}^{\mathsf{T}}
		\nonumber
		\\
	&= 
		\sum_{m_f}^{M_f}  \mathbf{F}_{m_f}^* 
		\left(\sum_{m_x}^{M_x} \mathbf{x}_{m_x}^* \mathbf{x}_{m_x}^{\mathsf{T}}\right) 
		\mathbf{F}_{m_f}^{\mathsf{T}},
		\nonumber
		\\
	&=
		\sum_{m_f}^{M_f}  \mathbf{F}_{m_f}^* 
		\mathbf{X}
		\mathbf{F}_{m_f}^{\mathsf{T}},
\end{align}
where $\mathbf{X} = \sum \mathbf{x}_{m_x}^* \mathbf{x}_{m_x}^{\mathsf{T}}$, the second equality holds if we use the same technique that is used for the transmitter and receiver separation, and $M_t = M_f M_x$ 
where $M_x$ and $M_f$ denote the number of pilot symbol vectors and precoders, respectively.
In the ideal case, it is desired that both $\mathbf{X}$ and $\sum \mathbf{F}_{m_f}^* \mathbf{F}_{m_f}^{\mathsf{T}}$ are to be the identity matrix to obtain $\sum \mathbf{s}_{m_t}^* \mathbf{s}_{m_t}^{\mathsf{T}} = \mathbf{I}$.  Then $M_x$ should be $L_t$, and accordingly $M_f$ becomes $\frac{N_t}{L_t}$. Since we do not want to lose beamforming capability, we control $M$ values by adjusting $M_x$. The value of $M_x$ determines the rank of $\mathbf{X}$. Then it boils down to the low rank approximation which can be expressed as
\begin{align}
	\min_{\mathbf{X}} \norm{ \mathbf{I} - \mathbf{X} }_F 
	\text{ subject to } \mathrm{rank}(\mathbf{X}) \leq M_x,
	\nonumber
\end{align}
and its well known analytic solution is to exploit truncated singular value decomposition (SVD). It implies that the smaller rank naturally leads $\mathbf{X}$ to have grater deviation from the identity matrix.  There exists infinite solutions; however, we have some criteria on choosing the solution. Firstly, all elements of all $\mathbf{x}_{m_x}$ should not be zero to take advantage of all possible beams. Secondly, all elements of $\mathbf{x}_{m_x}$ should have the identical magnitude to equally weight all beams. The $L_t$-point Discrete Fourier Transform (DFT) matrix satisfies these criteria, and any $M_x$ columns can be chosen from the DFT matrix and conclude the pilot codebook. The pilot codebook $\mathcal{X}$ can be expressed as
\begin{align}
	\mathcal{X} 
		&=
			\{
				\mathbf{x}_{m_x} : \forall m_x \in {1,2,\ldots,L_t}, \mathbf{U}_{L_t} = [\mathbf{x}_{1}, \mathbf{x}_{2}, \ldots, \mathbf{x}_{L_t} ]
			\} 
		\nonumber
		,
\end{align}
where $\mathbf{U}_{N} \in \mathbb{C}^{N \times N}$ is the $N$-point DFT matrix. According to $M_x$, $M$ is determined by $\frac{N_t}{L_t} \frac{N_r}{L_r} M_x$. 

We have assumed that $\sum \mathbf{F}_{m_t}^* \mathbf{F}_{m_t}^{\mathsf{T}}$ and $\sum \mathbf{W}_{m_r} \mathbf{W}_{m_r}^{\mathsf{H}}$ are the identity matrix. Taking into account the phase shifter based analog beamformers, $M_f = \frac{N_t}{L_t}$ and $M_r = \frac{N_r}{L_r}$, the column partition of the proper size DFT matrices can compose the beamformer codebooks. Namely then can be expressed as
\begin{align}
	\label{eq:FWcodebook}
	\mathcal{F} 
		&= 
			\left\{
			\mathbf{F}_{m_f} : \forall m_f \in \{1,2,\ldots, M_f=\frac{N_t}{L_t}\} ,
			\right.
			\nonumber \\
		& \quad \quad \quad \quad
			\left.
			\mathbf{U}_{N_t} = [\mathbf{F}_1, \mathbf{F}_2, \ldots, \mathbf{F}_{M_f}] 
			\right\}
			\nonumber , \\
	\mathcal{W}
		&=
			\left\{
			\mathbf{W}_{m_r} : \forall m_r \in \{1,2,\ldots, M_r=\frac{N_r}{L_r}\},
			\right.
			\nonumber \\
		& \quad \quad \quad \quad
			\left.
			\mathbf{U}_{N_r} = [\mathbf{W}_1, \mathbf{W}_2, \ldots, \mathbf{W}_{M_r}]
			\right\}
			,
\end{align}
where $\mathcal{F}$ and $\mathcal{W}$ denote the precoder and combiner codebooks, respectively. Another advantage of a DFT matrix is that the required phase shifter resolution in bits is $\log_2(\text{number of antennas})$. 

\begin{figure}[!t]
  \begin{algorithm}[H]
    \caption{Greedy Precoder Column Ordering}
    \label{alg:greedy_order}
    \begin{algorithmic}[1]
    \renewcommand{\algorithmicrequire}{\textbf{Input:}}
    \renewcommand{\algorithmicensure}{\textbf{Output:}}
    \Require $\mathbf{X}$, $\mathbf{U}$(=$N_t$-point DFT matrix)
    \Ensure $\mathbf{F}$
    \\ \textit{Initialization}: Set $\mathbf{F}$ to an empty matrix. 
    \For {$n=1$ to $N_t$ } \Comment{Choose $N_t$ columns}
    	\For {$m=1$ to the number of columns left in $\mathbf{U}$ } \\ \Comment{Go through a pool of possible columns}
    		\State $\mathbf{F}^{m} \leftarrow [\mathbf{F}, \mathbf{U}_{:,m}]$ 
    		\State $\mathbf{S}^{m} \leftarrow (\mathbf{F}^{m})^* (\mathbf{I} \otimes \mathbf{X}) (\mathbf{F}^{m})^{\mathsf{T}}$
    		\State Calculate $\mu(\mathbf{\Phi})_m$ with $\mathbf{S}^{m}$ by using \eqref{eq:mutcoh_Phi_simple}.
    	\EndFor
    	\State $m_{\min} = \argmin_m \mu(\mathbf{\Phi})_{m}$ 
    	\State $\mathbf{F} \leftarrow [\mathbf{F}, \mathbf{U}_{:,m_{\min}}]$ \Comment{Append the found vector to the codebook}
    	\State $\mathbf{U} \leftarrow \mathbf{U}_{:, [\ldots, m_{\min}-1, m_{\min}+1, \ldots]}$ \Comment{Remove the found vector from the pool}
    \EndFor \\
    \Return $\mathbf{F}$
    \end{algorithmic}
  \end{algorithm}
  \vspace*{-0.2in}
\end{figure}

\section{Precoder Column Permutation} 
\label{sec:precoder_column_permutation}
With $M_x < L_t$, $\mathbf{X}$ is not the identity matrix, and neither is \eqref{eq:sum_ss}. In this case, the column order in $\mathbf{F} = [\mathbf{F}_1, \mathbf{F}_2, \ldots, \mathbf{F}_{M_f}]$ and $\mathbf{x}_{m_t}$ selection result in changes not only in $\mathbf{S} = \sum \mathbf{s}_{m_t}^* \mathbf{s}_{m_t}^{\mathsf{T}}$, but also mutual coherence of $\mathbf{\Phi}$. In \cite{lee2016tcom}, the columns are proposed to be randomly permutated; however, we want to find a deterministic order so that the codebook can practically be used. Note that column permutation does not affect the fact that $\sum \mathbf{F}_{m_t}^{*} \mathbf{F}_{m_t}^{\mathsf{T}} = \mathbf{I}$. 

Since $\mathbf{F}$ has $N_t$ columns and $N_t$ is usually a large number, the exhaustive search for the best column order would not be feasible. With $N_t=64$, the number of permutations is $64! \approx 1.27\times10^{89} $. Therefore by adopting a greedy algorithm, we iteratively seek one column vector from the DFT matrix that achieves the lowest mutual coherence with the pre-selected vectors for the codebook. At the same time, $\mathbf{\Phi}$ is a $M L_t \times N_t N_r$ matrix, and calculating its mutual coherence directly from $\mathbf{\Phi}$ is computationally expensive. Thus we first obtain a simplified formulation for mutual coherence of the sensing matrix. Mutual coherence can also be defined as
\begin{align}
	\label{eq:mutcoh_Phi}
	\mu(\mathbf{\Phi}) = 
	\max_{i \neq j} \left| \left( \tilde{\mathbf{\Phi}}^{\mathsf{H}} \tilde{\mathbf{\Phi}} \right)_{ij} \right|,
\end{align}
where $\tilde{\mathbf{\Phi}}$ is the column-wise normalized version of $\mathbf{\Phi}$ which can also be expressed as $\tilde{\mathbf{\Phi}} = \mathbf{\Phi} \mathbf{\Sigma}^{-\frac{1}{2}}$ where $\mathbf{\Sigma} = \mathbf{I}_{N_t N_r} \circ (\mathbf{\Phi}^{\mathsf{H}} \mathbf{\Phi})$ and $\circ$ denotes the Hadamard product. Then we have
\begin{align}
	\tilde{\mathbf{\Phi}}^{\mathsf{H}} \tilde{\mathbf{\Phi}}
	&= 
		\left\{  
			\mathbf{I}_{N_t N_r} \circ \left(\mathbf{\Phi}^{\mathsf{H}} \mathbf{\Phi}\right)
		\right\}^{-\frac{1}{2}}
		\mathbf{\Phi}^{\mathsf{H}} \mathbf{\Phi}
		\left\{  
			\mathbf{I}_{N_t N_r} \circ \left(\mathbf{\Phi}^{\mathsf{H}} \mathbf{\Phi}\right)
		\right\}^{-\frac{1}{2}}
		\nonumber
		\\
	&=
		\left\{
			\mathbf{I}_{N_t N_r} \circ \left( 
					\mathbf{S} \otimes \mathbf{I}_{N_r}
				\right)
		\right\}^{-\frac{1}{2}}
		\left(\mathbf{S} \otimes \mathbf{I}_{N_r}\right)
		\times 
		\nonumber
		\\
	& \quad
		\left\{
			\mathbf{I}_{N_t N_r} \circ \left( 
					\mathbf{S} \otimes \mathbf{I}_{N_r}
				\right)
		\right\}^{-\frac{1}{2}}
		\nonumber
		\allowdisplaybreaks
		\\
	&=
		\left\{
			\left( \mathbf{S} \circ \mathbf{I}_{N_t} \right)^{-\frac{1}{2}} \otimes \mathbf{I}_{N_r}
		\right\}
		\left(\mathbf{S} \otimes \mathbf{I}_{N_r}\right) \times
		\nonumber
		\\
	& \quad
		\left\{
			\left( \mathbf{S} \circ \mathbf{I}_{N_t} \right)^{-\frac{1}{2}} \otimes \mathbf{I}_{N_r}
		\right\}
		\nonumber
		\allowdisplaybreaks
		\\
	&=
		\left\{
			\left( \mathbf{S} \circ \mathbf{I}_{N_t} \right)^{-\frac{1}{2}}
			\mathbf{S}
			\left( \mathbf{S} \circ \mathbf{I}_{N_t} \right)^{-\frac{1}{2}}
		\right\}
		\otimes \mathbf{I}_{N_r},
		\nonumber
\end{align}
where the second equality comes from \eqref{eq:ssWW} and \eqref{eq:FWcodebook}, and \eqref{eq:mutcoh_Phi} becomes
\begin{align}
	\label{eq:mutcoh_Phi_simple}
	\mu(\mathbf{\Phi}) = 
	\max_{i \neq j} \left| \left( 
		\left( \mathbf{S} \circ \mathbf{I}_{N_t} \right)^{-\frac{1}{2}}
		\mathbf{S}
		\left( \mathbf{S} \circ \mathbf{I}_{N_t} \right)^{-\frac{1}{2}} 
		\right)_{ij} \right|.
\end{align}
Here, $\mathbf{S}$ can be written in matrix form as $\mathbf{F}^*(\mathbf{I}_{\frac{N_t}{L_t}} \otimes \mathbf{X}) \mathbf{F}^{\mathsf{T}}$ by simplifying \eqref{eq:sum_ss}. In \eqref{eq:mutcoh_Phi_simple}, the matrix size is reduced by a factor of $N_r$ compared with \eqref{eq:mutcoh_Phi} in both dimensions. The matrix inversion and square root are taken on a diagonal matrix. 

For pilot codebook construction, $M_x$ column vectors are to be chosen from the $L_t$-point DFT matrix. As $L_t$ is usually not a large value, we perform exhaustive search for this selection. However, the first column of the DFT matrix should always be included in selection. This is because the elements in the first row of $\mathbf{F}_m$ are identical, and a sum of elements in the non-first column of the DFT matrix is zero. It leads to the first elements of $\mathbf{s}_{m} = \mathbf{F}_m \mathbf{x}_m$ being a zero and the first $N_r$ columns in $\mathbf{\Phi}_m$ zeros. Not having the first column of the DFT matrix as a pilot vector eventually ends up with the first $N_r$ columns in $\mathbf{\Phi}$ being zeros. It is also worth mentioning that the order of pilot vectors does not matter considering $\mathbf{X}$ is a sum of outer products of each pilot vector. 

\begin{figure}[t]
	\centering
	\subfloat[$M_x=2$]{
		\includegraphics[width=8.5cm]{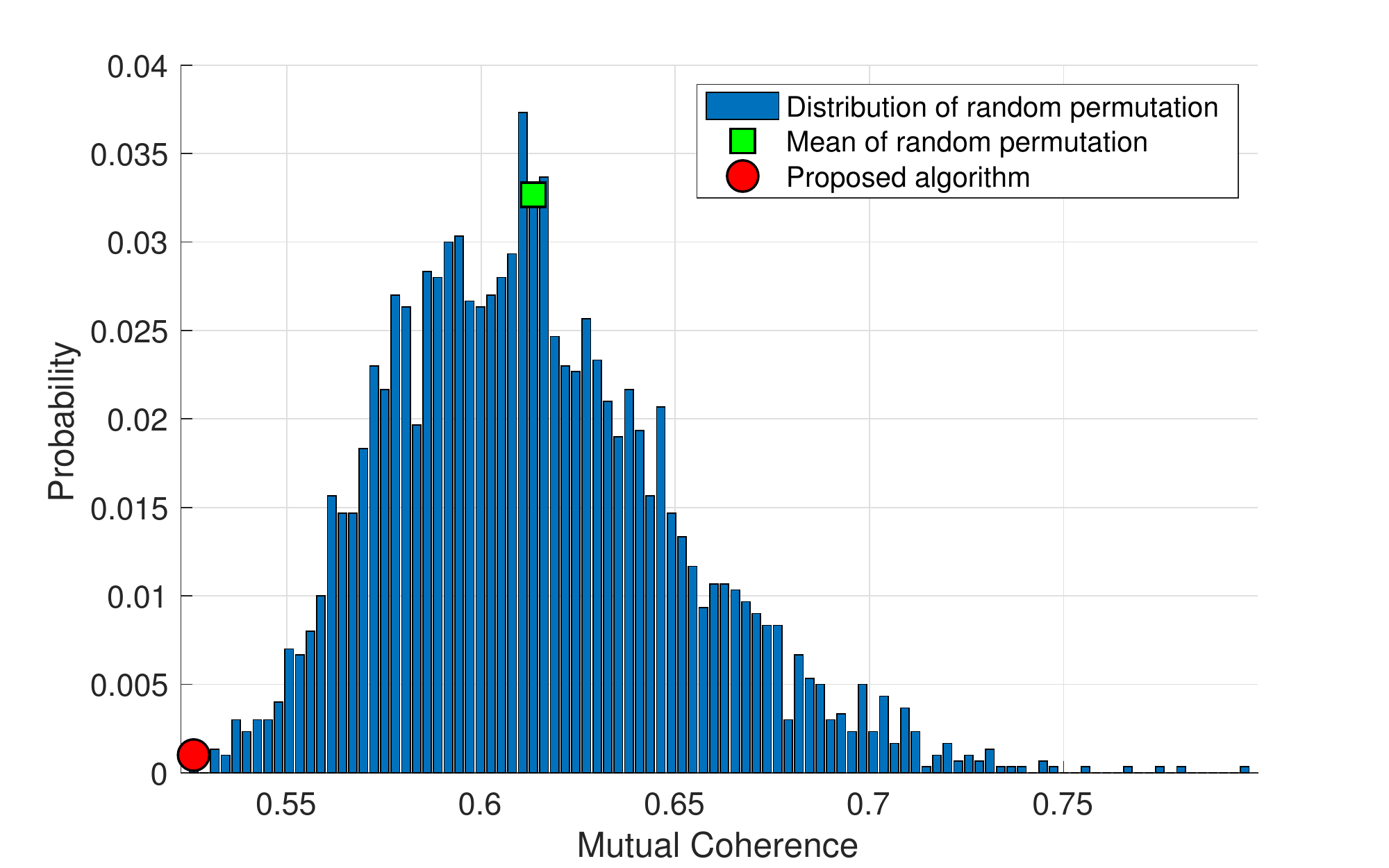}
		\label{fig:coh_dist_1}
	}
	\hfill
	\subfloat[$M_x=7$]{
		\includegraphics[width=8.5cm]{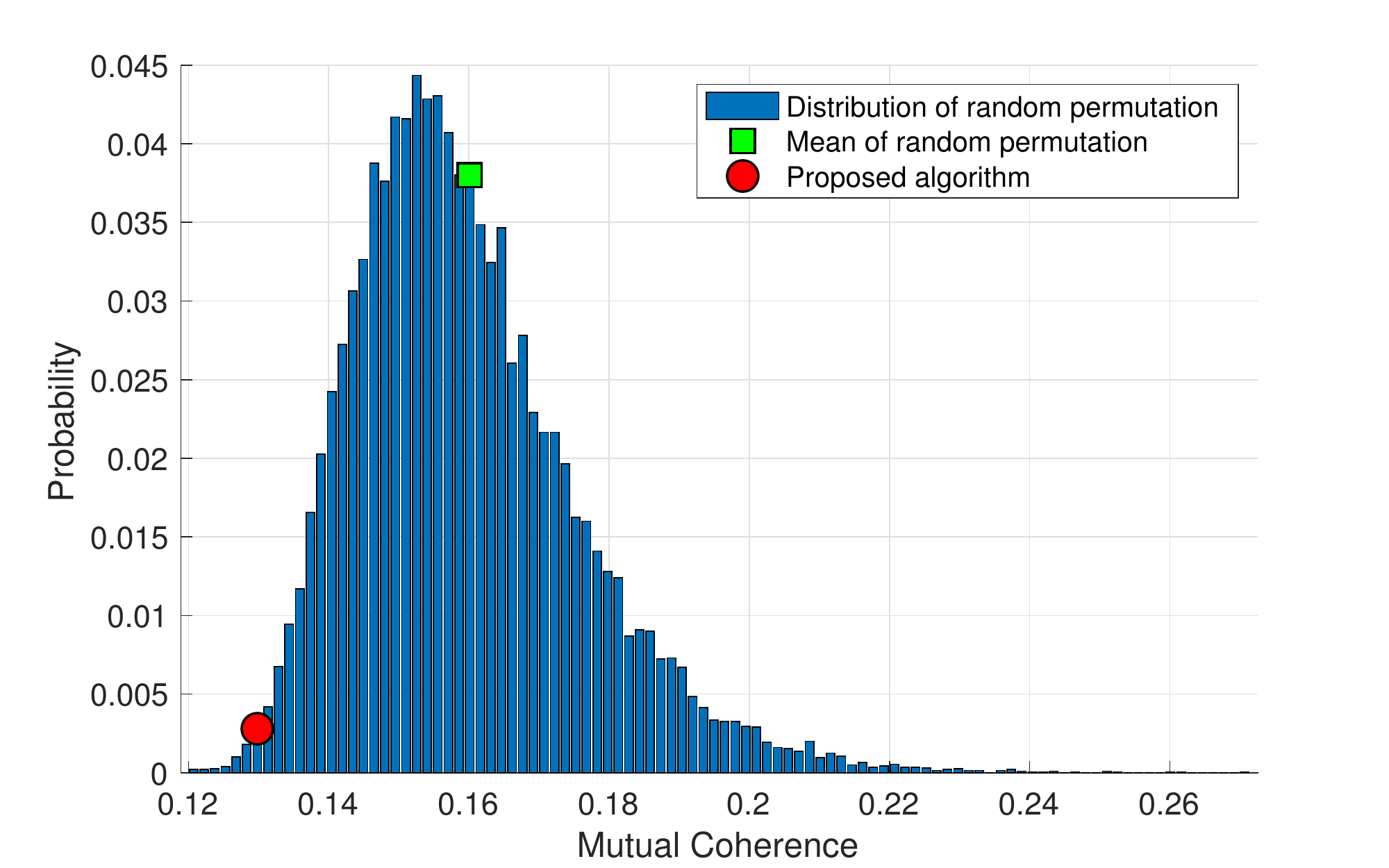}
		\label{fig:coh_dist_2}
	}
	\caption{Distribution of mutual coherence of the sensing matrix generated with 20,000 random precoder column permutation. The mean of the distribution and the mutual coherence of the proposed algorithm are marked as well. The parameters in this simulation is $N_t = 64$, $L_t=8$, and (a) $M_x=2$ and (b) $M_x=8$.}
	\label{fig:coh_dist}
	\vspace*{-0.15in}
\end{figure}

Taking into account all that is mentioned above, the greedy column ordering algorithm is given in Algorithm~\ref{alg:greedy_order}. $\mathbf{U}_{:,m}$ denotes the $m$-th column of the matrix $\mathbf{U}$. As Algorithm~\ref{alg:greedy_order} is for a given set of pilots, we perform the algorithm $\binom{L_t-1}{M_x-1}$ times to find the best pilot selection and precoder column order that achieves the lowest mutual coherence. 

The distribution of mutual coherence obtained by random column permutation and its mean value are provided in Fig.~\ref{fig:coh_dist} along with one obtained by the proposed algorithm. For illustration, two different values (two and seven) are considered for $M_x$. As shown in the figure for both cases, the mutual coherence of the proposed algorithm is lower than the mean of random permutation and is near the lowest value that random permutation can achieve. This observation can also be seen with all possible $M_x$ values, and the evaluated mutual coherence values from random permutation and the proposed algorithm are given in Table~\ref{tab:mutual_coherence}. For all possible $M_x$ values, the proposed algorithm achieves lower mutual coherence than the mean of mutual coherence distribution of random permutation, and as $M_x$ increases, mutual coherence declines. When $M_x = L_t (=8 \text{ in this case})$, the mutual coherence converges to zero. It implies that higher $M_x$ values make the CS algorithms perform more accurate channel estimation. 

\begin{table}[b]
	\caption{Mean coherence of random permutation and coherence of the proposed algorithm over various $M_x$ }
	\begin{center}
		\begin{tabular}{|c|c|c|c|c|c|c|c|c|}
			\hline
				$M_x$ & 1 & 2 & 3 & 4 & 5 & 6 & 7 & 8 \\
			\hline
				Permutation & 0.86 & 0.62 & 0.48 & 0.38 & 0.30 & 0.23 & 0.16 & 0\\
			\hline
				Proposed & 0.75 & 0.52 & 0.39 & 0.31 & 0.25 & 0.19 & 0.13 & 0 \\
			\hline
		\end{tabular}
		\label{tab:mutual_coherence}
	\end{center}
\end{table}

\section{Numerical Results} 
\label{sec:numerical_results}
In this section, we compare performance of the deterministic codebook obtained by the proposed algorithm with that of (i) the random phase shifter and pilot configuration, (ii) the MTC codebook with random precoder column permutation \cite{lee2016tcom} and (iii) the adaptive CS \cite{Alkh2014stsp}. For evaluation of the codebooks, OMP is employed except for the adaptive CS. The system parameters for simulation are as follows unless otherwise specified: $N_t=64$, $N_r=16$, $L_t=8$, $L_r=4$, $G_t = \gamma N_t$, $G_r = \gamma N_r$, $b_{PS}=6$ and $N_p=4$
where $\gamma=1.5 (>1)$ is the grid multiplier. 
For the adaptive CS, $G_t = G_r = 96$, which results in $M=864$. The normalized mean squared error (NMSE) is defined as $\mathbb{E}[ \lVert \mathbf{H} - \hat{\mathbf{H}} \rVert_F^2 / \lVert \mathbf{H} \rVert_F^2]$ where $\mathbf{H}$ and $\hat{\mathbf{H}}$ are the true and the estimated channel matrix, respectively, 
SNR is defined as $\rho / \sigma^2$, 
and $(\cdot)_{F}$ denotes the Frobenius norm of a matrix. 
Multipath components of the channels, in simulations, have AoDs and AoAs that are not necessarily aligned with the grids of the dictionary. The complete source code is available \cite{projectcode}.

\begin{figure}[t]
	\centering
	\includegraphics[width=9cm]{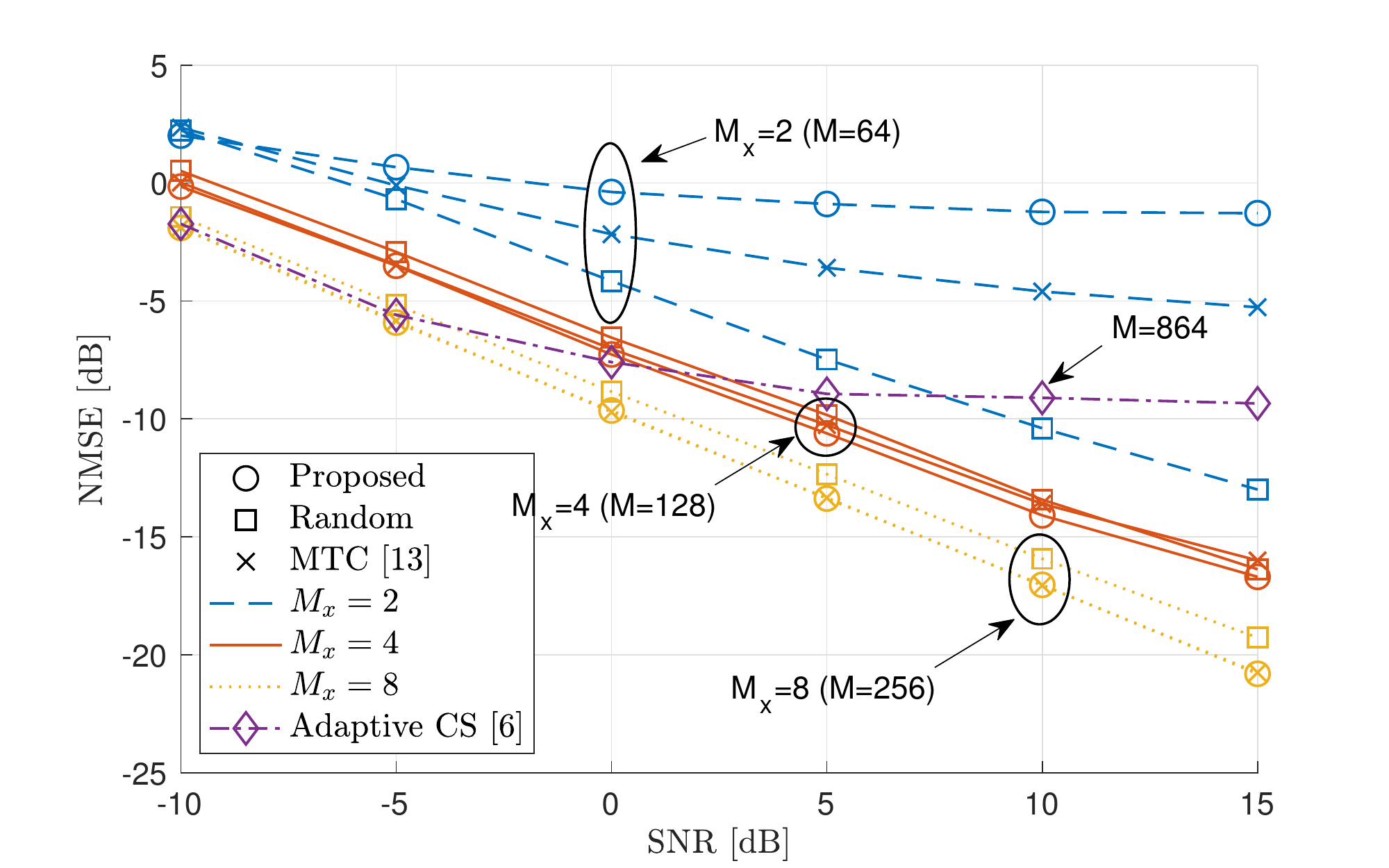}
	\caption{NMSE vs. SNR with the proposed, random and MTC codebooks using OMP \cite{lee2016tcom} along with the adaptive CS channel estimation \cite{Alkh2014stsp}.}
	\label{fig:nmse1}
	\vspace*{-0.15in}
\end{figure}

Fig.~\ref{fig:nmse1} shows NMSE as a function of the transmit SNR, and the three codebooks and the adaptive CS are considered. The first observation is that NMSE decreases with SNR and the number of snapshots ($M$). With an increase in $M$, the estimator can take more measurements, which improves performance of CS-based estimator. At the same time, however, higher $M$ also requires a longer measurement time and more computation. Thus we have a performance versus time and power consumption trade-off. 

Three values of $M (=64, 128, 256)$ are considered in Fig.~\ref{fig:nmse1}, and performance ranking of the codebooks changes depending on the value. The proposed codebook achieves the lowest NMSE among the three when $M$ is either $128$ or $256$. With $M=256$, its performance is the best together with the MTC, and $M=64$ makes it the worst. It implies that the proposed algorithm works well when the number of snapshots is not too small. This is due to the fact that $\mathbf{X}$ deviates from the identity matrix with small $M_x$. Different approaches may be explored to address this issue, which will be our future work. 

The adaptive CS generally needs a large number of snapshots which is determined by the grid size. Even with more snapshots, estimation errors are greater than the considered codebooks including the proposed one. In Fig.~\ref{fig:nmse1}, the adaptive CS has $M=864$, but the NMSE is higher than the proposed codebook with $M=256$ across the SNR range and than that with $M=128$ in medium and high SNR regimes.  

\begin{figure}[t]
	\centering
	\subfloat[$\text{SNR}=15$ dB]{
		\includegraphics[width=9.3cm]{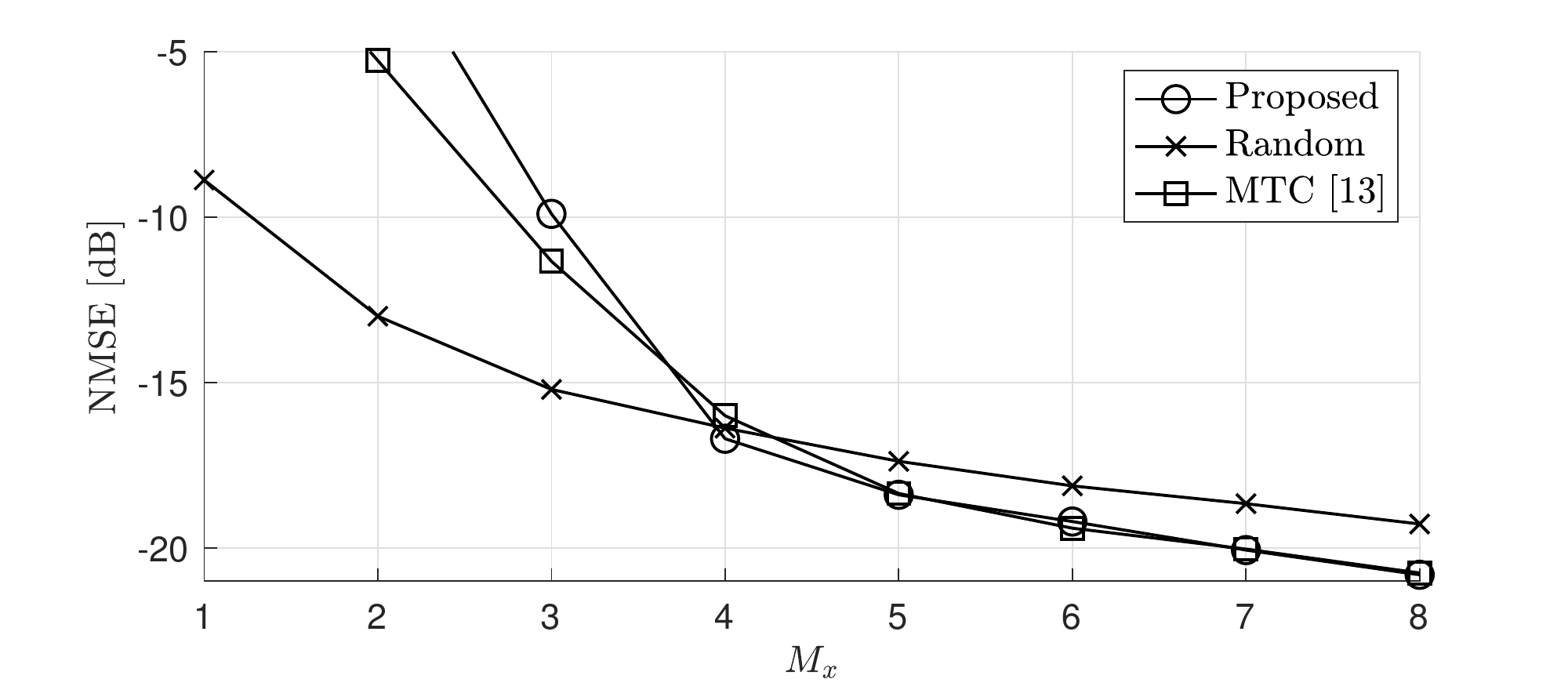}
		\label{fig:nmse2_0}
	}
	\hfill
	\subfloat[$\text{SNR}=0$ dB]{
	\includegraphics[width=9.3cm]{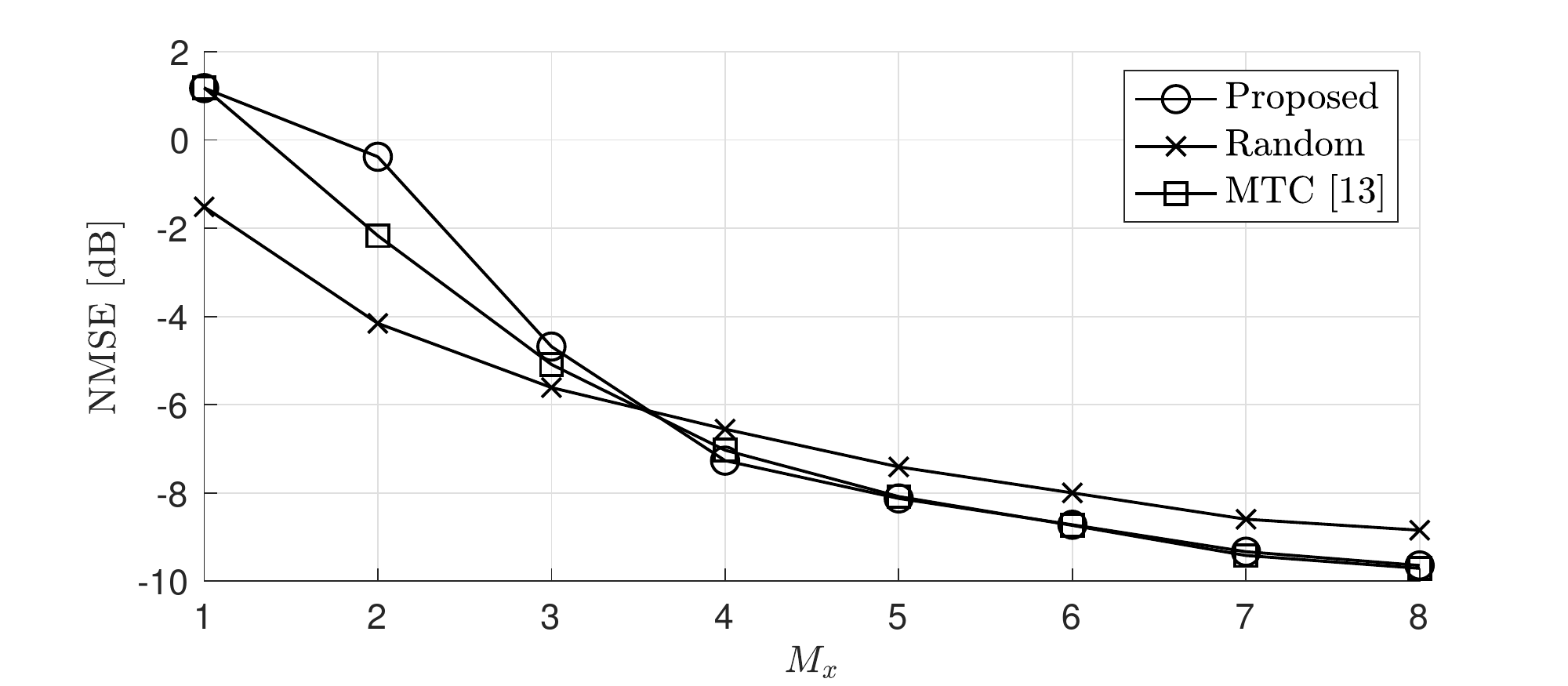}
	\label{fig:nmse2_1}
	}
	\caption{NMSE vs. $M_x$ with SNR of (a) $15$ dB and (b) $0$ dB.}
	\label{fig:nmse2}
	\vspace*{-0.15in}
\end{figure}

To illustrate the relationship between performance and $M_x$, Fig.~\ref{fig:nmse2} is provided where SNR is fixed at either $15$ or $0$ dB. The NMSE of the random codebook starts from a relatively low NMSE and gradually declines with $M_x$. The other two codebooks, on the other hand, start from high NMSE, drop at low $M_x$, and gradually decline in medium and high $M_x$ regimes. Due to the steep drops in the low $M_x$ regime, NMSE curves of the proposed and MTC codebooks cross that of random codebook. For $15$ dB SNR, the crossover happens at $M_x=4$ with the proposed codebook which is earlier than $M_x=5$ with the MTC codebook. For $0$ dB SNR, the it happens at $M_x=4$ for both codebooks. 
With system configurations such as one used in this simulation, thus, the proposed codebook is preferred when $M_x$ is greater than three as (i) it achieves NMSE that is lower or equal to that of the MTC codebook and (ii) is lower than that of the random codebook when $M_x>3$ and (iii) the proposed codebook has no randomness. 

\begin{figure}[t]
	\centerline{\includegraphics[width=9.3cm]{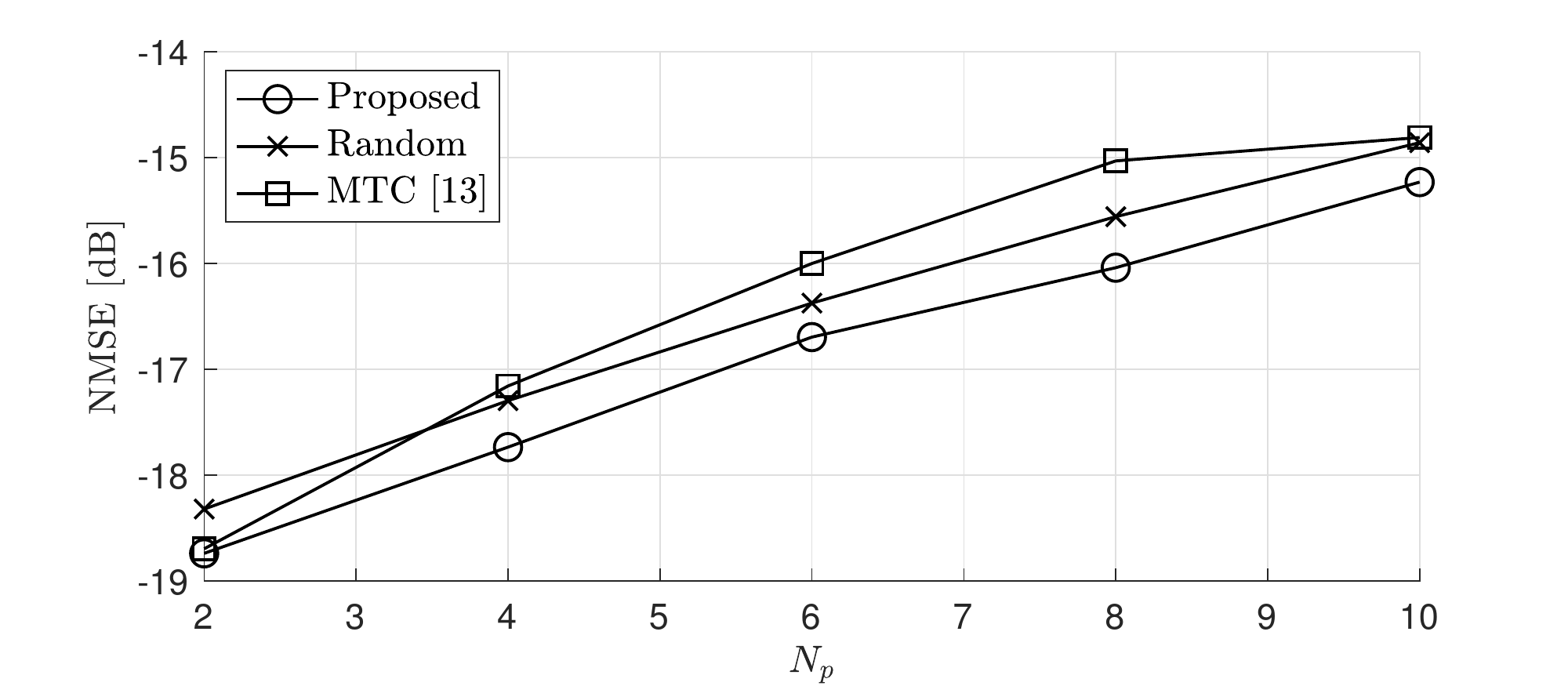}}
	\caption{NMSE vs. $Np$ channel paths with 15 dB SNR and $M_x=4$}
	\label{fig:nmse3}
\end{figure}

The number of channel paths $N_p$ affects channel estimation performance as well since it is directly related to the sparsity of the channel vector being estimated. In Fig.~\ref{fig:nmse3}, SNR and $M_x$ are fixed at 15 dB and four, respectively. In this figure, the proposed codebook yields the lowest estimation error across the $N_p$ values. The order of the MTC and the random codebooks varies depending on $N_p$. Since $N_p$ varies and is determined by the channel environment, the proposed codebook would be a good choice as it achieves the lowest channel estimation error regardless of the number of channel paths. 

\section{Conclusion} 
\label{sec:conclusion}
In this paper, we proposed a codebook design method for mmWave channel estimation in hybrid beamforming communication systems based on mutual coherence minimization. We first obtained the criteria for codebooks of pilots, precoders and combiners, and proposed the pilot selection and precoder column ordering algorithm for further mutual coherence reduction. The mutual coherence distribution of random column permutation was provided to show the proposed greedy precoder column ordering algorithm achieves lower mutual coherence than random permutation. 

We also provided the channel estimation simulation results using OMP for performance comparison between the proposed, the random and the MTC codebooks. 
The proposed method provides the best tradeoff between channel estimation performance and measurement time, and has more accurate channel estimation vs.	 number of channel paths.
\bibliographystyle{IEEEtran}
\bibliography{bib/paper}
\end{document}